\documentclass[aps,prb,twocolumn,showpacs,superscriptaddress]{revtex4}
\usepackage{graphicx}
\usepackage{dcolumn}
\usepackage{bm}
\usepackage{amssymb}
\usepackage{amsmath}
\usepackage{epsfig}

\begin{document}
\title{Parametric four-wave mixing toolbox for superconducting resonators}
\author{A. V. Sharypov, Xiuhao Deng, and Lin Tian}
\email{LTian@ucmerced.edu}
\affiliation{University of California, Merced, 5200 North Lake Road, Merced, California 95343, USA}

\begin{abstract}
We study a superconducting circuit that can act as a toolbox to generate various Bogoliubov-linear and nonlinear quantum operations on the microwave photon modes of superconducting resonators within one single circuit. The quantum operations are generated by exploring dispersive four-wave mixing (FWM) processes involving the resonator modes. Different FWM geometries can be realized by adjusting the circuit parameters and by applying appropriate microwave drivings. We illustrate this scheme using a circuit made of two superconducting qubits that couple with each other. Each qubit couples with one superconducting resonator. We also discuss main sources of quantum errors in this system and study the fidelity of the quantum operations by numerical simulation. Our scheme provides a practical approach to realize quantum information protocols on superconducting resonators.
\end{abstract}
\date{\today}
\maketitle

\section{introduction}
Superconducting quantum circuits have been intensively studied as building blocks for quantum information processing.\cite{SchonReview, ClarkeReview, BylanderOliverNPhys2011, deGrootMooijNPhys2010, DeppeGrossNPhys2008, MariantoniMartinisScience2011, AstafievScience2010, BishopSchoelkopfNPhys2009, DiCarloNature2010Entangle3Qbit, BozyigitWallraffNPhys2010, JohnsonNPhys2010, BergealDevoretNPhys2010, MalletEsteveNPhys2009} Among these devices, superconducting microwave resonators have demonstrated relatively high quality factors and strong coupling with various superconducting qubits.\cite{SchoelkopfReview} Various quantum optical phenomena and novel quantum many-body effects involving microwave photons, in particular, circuit quantum electrodynamics (CQED), have been observed in the superconducting resonators.\cite{NoriArtAtom} 

The generation of non-classical states such as Fock states, entangled states, and NOON states in the superconducting resonators has been studied in recent experiments and theoretical proposals.\cite{HofheinzNature2009, MariantonilelandNPhys2011, Zakka-BajjaniAumentadoNPhys2011ColorState, MerkelWilhelmNJP2010, HWangClelandPRL2011, AldanaBruderPRB2011, MYChenPRB2009, TianPRL2011EntangledState, PBLiPRA2011SqueezedState, MariantoniPRB2008, ReutherSolanoHanggiPRB2010, JQLiaoPRA2009, StrauchSimmondsPRL2010, ZagoskinNoriPRL2009}  In most of these schemes, quantum state manipulation is achieved via the coupling between superconducting resonators and superconducting qubits.  Several schemes have been proposed to generate unitary transformations on the microwave photons by engineering effective interaction Hamiltonians between the resonator modes, including Kerr and cross-Kerr interactions, beam-splitter operation, and squeezing operation.\cite{ChirollioPRL2010, KumarPRA2010, RebicMilburnPRL2009GiantKerrCQED, CDengLupascu, TianNJP2008}. In these schemes, the proposed circuits can only realize specific quantum operation on the resonator modes. In a recent work, Langford and coauthors studied an interesting idea of generating quantum operations starting from a nonlinear-Kerr interaction between different optical (or microwave) modes. By applying selected classical pumps, full set of quantum operations can be generated between selected modes.\cite{LangfordNature2011}

\begin{figure}
\includegraphics[width=8cm,clip]{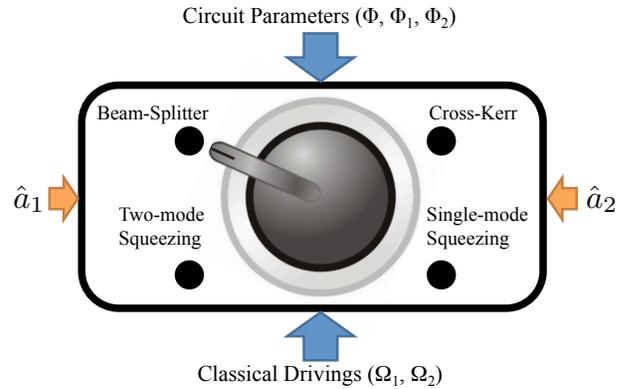}
\caption{Schematic illustration of the dispersive FWM toolbox with a dial plate. Two resonator modes $\hat{a}_{1}$ and $\hat{a}_{2}$ are connected to the toolbox. The toolbox is made of a superconducting circuit with controllable parameters $(\Phi,\Phi_{1},\Phi_{2})$ and classical drivings with Rabi frequencies $(\Omega_{1}, \Omega_{2})$.  By adjusting the parameters and drivings to appropriate values, the toolbox is switched on to perform selected quantum operation on the resonator modes as labelled on the dial plate.}
\label{fig1_toolbox}
\end{figure}
All of the quantum operations that are crucial for implementing quantum information protocols on resonator modes can be constructed from a basic set of quantum operations. For both discrete-state and continuous variable quantum information protocols on the resonator modes, these basic operations include the Bogoliubov-linear operations such as the beam-splitter operation, the squeezing operation and the phase shifter, and nonlinear operations such as the cross-Kerr interaction.\cite{KokRMP2007, LloydBraunsteinPRL1998} A discussion of these operations can be found in Appendix A. In this work, we present a scheme that can generate all of these basic operations using one single circuit. This toolbox  is made of two superconducting qubits coupled with each other to form a quantum four-level system. Each qubit interacts with one superconducting resonator. By adjusting the parameters of the toolbox, we design dispersive four-wave mixing (FWM) processes to generate effective quantum operations on the resonator modes, as is illustrated in Fig.~\ref{fig1_toolbox}. During the operations, the quantum four-level system is always preserved in its quantum ground state by large detunings, and hence this scheme is a parametric scheme. Using numerical simulation, we show that high-fidelity quantum operations can be achieved with realistic circuit parameters. In particular, we demonstrate the realization of a controlled phase gate using the proposed scheme and show that fidelity higher than $99\%$ can be achieved. Compared with previous schemes,\cite{ChirollioPRL2010, KumarPRA2010, RebicMilburnPRL2009GiantKerrCQED, CDengLupascu, TianNJP2008} our proposal provides a switchable circuit that can generate all of the basic quantum operations by adjusting the parameters of the toolbox. Hence, the proposed scheme can advance the scalability of quantum information protocols on superconducting resonators by connecting a network of resonators with such toolboxes. 

The paper is organized as the following. In Sec.~\ref{sec2}, we present the general idea of the toolbox and illustrate the idea with a specific circuit made of two coupled superconducting charge qubits. In Sec.~\ref{sec3}, we describe the dispersive FWM processes for generating effective Hamiltonians on the resonator modes using the toolbox. The realizations of four quantum operations are presented in detail in Sec.~\ref{sec4}. In Sec.~\ref{sec5}, we discuss the main sources of quantum errors in this scheme and present our numerical simulation of the controlled phase gate using the effective cross-Kerr interaction. Conclusions are given in Sec.~\ref{sec6}. In Appendix A, we briefly discuss all of the basic quantum operations and their roles in the quantum information processing for the resonator modes.

\section{circuit\label{sec2}}
The central element of the toolbox is a quantum four-level system that couples with the superconducting resonators and can be constructed in many ways. For the discussion in this paper, we will consider a circuit for the toolbox that is made of two superconducting qubits coupling with each other. The Hamiltonian for the total system has the form
\begin{equation}
H_{tot}=H_{q}+H_{r}+H_{p},\label{Hexact}%
\end{equation}
which includes the Hamiltonian for the four-level system $H_{q}=\sum_{j}E_{j}|j\rangle\langle j|$ with eigenstates $|j\rangle$ and eigenenergies $E_{j}$ ($j=a,\,b,\,c,\,d$), the Hamiltonian for the resonators and their couplings to the qubits 
\begin{equation}
H_{r}=\sum_{i}\hbar\omega_{a_{i}}\hat{a}_{i}^{\dagger}\hat{a}_{i}+\hbar g_{i}\sigma_{xi}(\hat{a}_{i}^{\dagger}+\hat{a}_{i}),\label{Hr}
\end{equation}
with resonator frequencies $\omega_{a_{i}}$ and coupling constants $g_{i}$, and the Hamiltonian for the classical drivings on the qubits $H_{p}=\sum \hbar\Omega_{i}(t)\sigma_{xi}$ with driving amplitudes $\Omega_{i}$. Without loss of generality, we  have assumed that each resonator couples only with one qubit via the $\sigma_{xi}$ operator and the classical driving is also on the $\sigma_{xi}$ operator. Other forms of coupling and driving could also be considered for the toolbox. 

The coupling terms in $H_{r}$ (the driving terms in $H_{p}$) generate transitions between the eigenstates of the four-level system which involve absorption or emission of resonator photons (classical field). The transition matrix elements induced by these terms can be derived by projecting the $\sigma_{xi}$ operator in the eigenbasis $|j\rangle$.  

Superconducting qubits in various parameter regimes and circuit geometries have been studied, including flux qubit, phase qubit, charge qubit, transmon qubit, and \emph{etc.}.\cite{SchonReview, ClarkeReview} The qubits can be controlled by external electromagnetic fields such as the magnetic flux in the loop of a superconducting quantum interference device (SQUID) and the bias voltage on a superconducting island, depending on specific circuit design. Different coupling mechanisms between superconducting qubits have also been studied, such as capacitive coupling, Josephson coupling, and inductive coupling.\cite{PintoPRB2010, YXLiuPRL2006QubitCoupling, HarrisPRL2007, IlÕichevPRB2010CoupledFluxQubits, SrinivasanHouckPRL2011TunableCupling} Decoherence in superconducting qubits has improved greatly in the past few years with $T_{2}^{\star}$ exceeding $95\,\mu\textrm{s}$ recently observed.\cite{KochSchoelkopfPRA2007TransmonQubit, HPaikSchoelkopfPRL2011, ZKimWellstoodPalmerPRL2011, RigettiSteffenPreprint2012}

\begin{figure}
\includegraphics[width=8.5cm,clip]{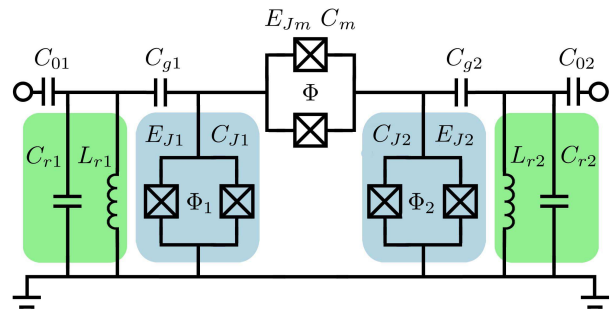}
\caption{Schematic circuit for the toolbox. The resonators are labelled by capacitances $C_{ri}$ and inductances $L_{ri}$.}
\label{fig2_charge}
\end{figure}
To illustrate our scheme, we study a toolbox made of two superconducting charge qubits.\cite{SchonReview} We want to emphasize, however, the FWM approach studied here is a general scheme that can be applied to other superconducting qubits with different forms of coupling. As is shown in Fig.~\ref{fig2_charge}, the charge qubits couple with each other via a tunable Josephson junction where the effective Josephson energy $E_{Jm}$ can be adjusted by varying the magnetic flux $\Phi$ in the loop.  The Josephson energies $E_{Ji}$ of the qubit junctions can be adjusted by changing the magnetic flux $\Phi_{1,2}$ in the qubit loops. The Hamiltonian for the coupled qubits can be derived using a Lagrangian approach with
\begin{eqnarray}
H_{q}&= &(E_{J1}/2)\sigma_{z1}+(E_{J2}/2)\sigma_{z2}+H_{int} \label{Hq}\\
H_{int}&= &E_{mx}\left(  \sigma_{x1}\sigma_{x2}+b_{0}\sigma_{y1}\sigma_{y2}+b_{0}\sigma_{z1}\sigma_{z2}\right) \label{Hint}
\end{eqnarray}
where $E_{mx}$ is the charging energy of the capacitance $C_{m}$ of the coupling junction and $b_{0}=E_{Jm}/4E_{mx}$ is the ratio between the Josephson energy and the charging energy. Note that the coupling junction is a larger junction with its Josephson energy larger than its charging energy, in contrast to the junctions of the charge qubits. Here, we assume the qubits are biased to have zero charging energy so that the qubit energies are the Josephson energies $E_{Ji}$.\cite{DengarXiv} We define the total capacitance connected to the superconducting island of the $i$th qubit as $C_{\Sigma i}=C_{Ji}+C_{gi}+C_{m}$ and the total capacitance connected to the $i$th resonator as $C_{\Sigma ri}=C_{ri}+C_{gi}+C_{0i}$, where $C_{gi}$ is the capacitance that couples a resonator to its corresponding qubit, $C_{Ji}$ is the Josephson capacitance, $C_{ri}$ is the resonator capacitance, and $C_{0i}$ is the capacitance that couples to external circuit, as labeled in Fig.~\ref{fig2_charge}. With the capacitances satisfying $C_{\Sigma ri}\gg C_{\Sigma i}\gg C_{m}$, the capacitive coupling $E_{mx}$ can be derived as
\begin{equation}
E_{mx}\approx C_{m}e^{2}/\left[C_{\Sigma 1}C_{\Sigma 2}- C_{m}^{2}\right].\label{Emx}
\end{equation}
The eigenenergies of the Hamiltonian $H_{q}$ are
\begin{subequations}
\begin{align}
E_{a} &  =E_{mx}b_{0}-E_{s+}/2,\label{Energy1}\\
E_{b} &  =-E_{mx}b_{0}-E_{s-}/2,\label{Energy2}\\
E_{c} &  =-E_{mx}b_{0}+E_{s-}/2,\label{Energy3}\\
E_{d} &  =E_{mx}b_{0}+E_{s+}/2,\label{Energy4}
\end{align}
\end{subequations}
where $E_{s\pm}=\sqrt{(E_{J1}\pm E_{J2})^{2}+4E_{mx}^{2}( 1\mp b_{0})^{2}}$ can be adjusted by varying the qubit energies $E_{Ji}$ and the Josephson coupling $E_{Jm}$. The eigenstates are
\begin{subequations}
\begin{align}
\left\vert a\right\rangle  &  =- \sin\theta_{+}\left\vert 0_{1}0_{2}\right\rangle +\cos\theta_{+}\left\vert 1_{1}1_{2}\right\rangle, \label{State1}\\
\left\vert b\right\rangle  &  =\cos\theta_{-}\left\vert 0_{1}1_{2}\right\rangle -\sin\theta_{-}\left\vert 1_{1}0_{2}\right\rangle, \label{State2}\\
\left\vert c\right\rangle  &  =\sin\theta_{-}\left\vert 0_{1}1_{2}\right\rangle +\cos\theta_{-}\left\vert 1_{1}0_{2}\right\rangle, \label{State3} \\ 
\left\vert d\right\rangle  &  =\cos\theta_{+}\left\vert 0_{1}0_{2}\right\rangle+\sin\theta_{+}\left\vert 1_{1}1_{2}\right\rangle, \label{State4}
\end{align}
\end{subequations}
where $|0_{i}\rangle$ and $|1_{i}\rangle$ are the single-qubit eigenstates in the $\sigma_{zi}$ basis and $\sin\theta_{\pm} =\sqrt{\left[E_{s\pm}\pm\left(  E_{J1}\pm E_{J2}\right) \right]/2E_{s\pm}}$.

The resonator-qubit coupling $g_{i}$ for the $i$th resonator in Eq.~(\ref{Hr}) can also be derived using the Lagrangian approach with
\begin{equation}
g_{i}=\left(  \frac{C_{gi}C_{\Sigma \bar{i}}}{C_{\Sigma 1}C_{\Sigma 2}-C_{m}^{2}}\right)  \left(\frac{e^{2}}{2C_{\Sigma ri}}\hbar\omega_{a_{i}}\right)^{1/2} \label{coupling}\end{equation}
where $C_{\Sigma \bar{i}}$ is the total capacitance of the $\bar{i}$th qubit with the index $\bar{i}$ referring to the qubit in the opposite side of the toolbox from the $i$th resonator. We also find that the $i$th resonator couples with the $\bar{i}$th qubit as well as the $\bar{i}$th resonator in the opposite side of the toolbox due to the cross-talk of the circuit elements. These indirect couplings have the forms of $\hbar g_{i}^{(2)}\sigma_{xi}(\hat{a}^{\dagger}_{\bar{i}}+\hat{a}_{\bar{i}})$ and $\hbar g^{(3)}(\hat{a}_{1}+\hat{a}_{1}^{\dag})(\hat{a}_{2}+\hat{a}_{2}^{\dag})$ with coupling constants much weaker than the dominant couplings $g_{i}$ in Eq.~(\ref{coupling}), as will be discussed in Sec.~\ref{sec5}.

Both the resonator-qubit coupling and the classical driving $H_{p}$ are associated with the qubit operators $\sigma_{xi}$. In the eigenbasis in Eqs.~(\ref{State1}, \ref{State2}, \ref{State3}, \ref{State4}), we have
\begin{subequations}
\begin{align}
\sigma_{x1} & =  \cos(\theta_{+}-\theta_{-})(\sigma_{ab}+\sigma_{dc}) \nonumber  \\ 
& + \sin(\theta_{+}-\theta_{-})(\sigma_{db}-\sigma_{ac}) +h.c.\label{sigmaX1}\\ 
\sigma_{x2}  & =  -\sin(\theta_{+}+\theta_{-})(\sigma_{ab}-\sigma_{dc})  \nonumber \\ 
& +  \cos(\theta_{+}+\theta_{-})(\sigma_{ac}+\sigma_{db}) +h.c.\label{sigmaX2}%
\end{align}
\end{subequations}
where $\sigma_{ij}=\left\vert i\right\rangle \left\langle j\right\vert $ defines the transition operator connecting the eigenstates $|i\rangle$ and $|j\rangle$. From these expressions, it follows that each resonator (and classical driving) generally couples to four transitions. For example, the $\sigma_{x1}$ term includes the $\sigma_{ab}, \sigma_{dc}$ transitions with amplitude $g_{1} \cos(\theta_{+}-\theta_{-})$ and the $\sigma_{ac}, \sigma_{db}$ transitions with amplitude $\pm g_{1}\sin(\theta_{+}-\theta_{-})$.

In the following sections, we will show how to engineer the energy levels of the quantum four-level system to have the resonators couple only with selected transitions by adjusting the circuit parameters.

\section{dispersive FWM scheme\label{sec3}}
The effective Hamiltonians to implement quantum operations on the resonators can be realized via the resonator-qubit coupling and classical driving in the circuit studied above. Here, we exploit four-photon processes,\cite{BoydBook, BiPhotonsSharypov, SchmidtImamogluOptLett1996GiantKerrByEIT, ImamogluPRL1997PhotonBlockadeEIT, BrandaoNJP2008Nonlinearity, HartmannPlenioPRL2007} in which single-photon transitions and two-photon processes are in the dispersive regime with large detunings while the designated four-photon processes are nearly at resonance. As an example, in Fig.~\ref{fig4_FWM}a, consider a classical driving of frequency $\omega_{1}$ generating the transition $\sigma_{ac}$ with Rabi frequency $\Omega_{1}$ and the resonator mode $\hat{a}_{1}$ generating the transition $\sigma_{dc}$ with effective transition matrix element $\tilde{g}_{1}$. For the single-photon transition induced by the classical driving, the dispersive condition requires that $|\Delta_{1}|=|\omega_{1}-E_{ca}/\hbar|\gg\Omega_{1}$; and for the single-photon transition induced by mode $\hat{a}_{1}$, it requires that $|\omega_{a_{1}}-E_{dc}/\hbar| \gg \tilde{g}_{1}\sqrt{n_{1}}$ with $E_{ij}=E_{i}-E_{j}$ and $n_{1}$ being the average photon number in $\hat{a}_{1}$. For the two photon process involving these two transitions, it requires that $|\Delta_{1}\delta|\gg \Omega_{1}\tilde{g}_{1}\sqrt{n_{1}}$ with $|\delta|=|\omega_{1}+\omega_{a_{1}}-E_{da}/\hbar|$ being the two-photon detuning. Under the dispersive conditions, the dominant physical processes in this scheme are four-photon processes which can generate effective coupling between the resonator modes. Because the dispersive conditions prevent real transitions, the quantum four-level system is preserved in its ground state during the operation. The processes studied here are hence parametric schemes where the four-level system is subject to a ``quantum" energy shift.\cite{BoydBook} 

\begin{figure}
\includegraphics[width=7.5cm,clip]{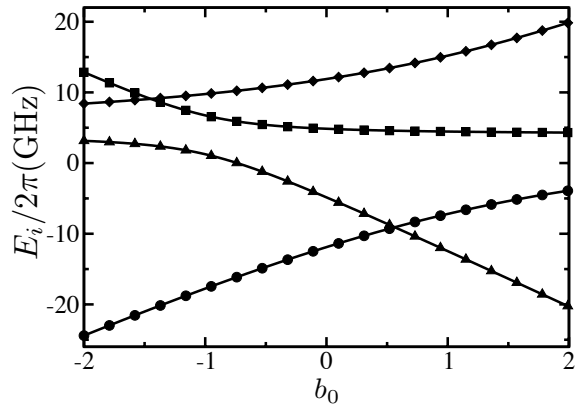}%
\caption{Eigenenergies of the quantum four-level system with circles for $E_{a}$, up-triangles for $E_{b}$, squares for $E_{c}$, and diamonds for $E_{d}$. The parameters are
$E_{J1}/2\pi\hbar=8.45\,\textrm{GHz}$, $E_{J2}/2\pi\hbar=13.95\,\textrm{GHz}$, and $E_{mx}/2\pi\hbar=4\,\textrm{GHz}$ (same parameters as used in the cross-Kerr operation in Sec.~\ref{crossKerr}). }%
\label{fig3_energy}%
\end{figure}
To implement a quantum operation, we need to adjust the parameters of the superconducting circuit to find appropriate effective coupling constants and energy separations between the eigenstates. For the circuit in Fig.~\ref{fig2_charge}, we adopt two approaches to determine the circuit parameters. The first approach uses the relation in Eq.~(\ref{sigmaX1}, \ref{sigmaX2}) to tune the effective transition matrix elements by adjusting the angles $\theta _{\pm}$. By adjusting the circuit parameters $E_{J1}, E_{J2}, b_{0}$, the angles can be varied in a large range. The second approach exploits the controllability of the energy levels to engineer large detunings to suppress unwanted transitions. As an example, in Fig.~\ref{fig3_energy}, we plot the eigenenergies as functions of the ratio $b_{0}$ (defined in Sec.~\ref{sec2}). As $b_{0}$ varies, the state $|a\rangle $ ($|c\rangle $) can be either above or below the state $|b\rangle $ ($|d\rangle $), which provides the possibility to arrange both the order and the energy separation of the eigenstates. To make, e.g. the mode $\hat{a}_{1}$ only couples strongly to the $\sigma _{ab}$ transition, we choose the parameters to have: (1) $\cos(\theta _{+}-\theta _{-})\gg \sin(\theta_{+}-\theta _{-})$ and (2) $|\omega _{a_{1}}-E_{ba}/\hbar |\ll |\omega _{a_{1}}-E_{dc}/\hbar|$. The first condition significantly reduces the effective coupling between $\hat{a}_{1}$ and the $\sigma _{ac}, \sigma _{db}$ transitions and the second condition suppresses the $\hat{a}_{1}\sigma _{dc}$ term by large detuning. Now, mode $\hat{a}_{1}$ couples mainly with the $\sigma _{ab}$ transition. In general, by combining the above two approaches, all quantum operations described in Appendix A can be realized with the FWM scheme within one single circuit.

\section{realization of operations\label{sec4}}
\begin{figure}[ptb]
\includegraphics[width=7cm,clip]{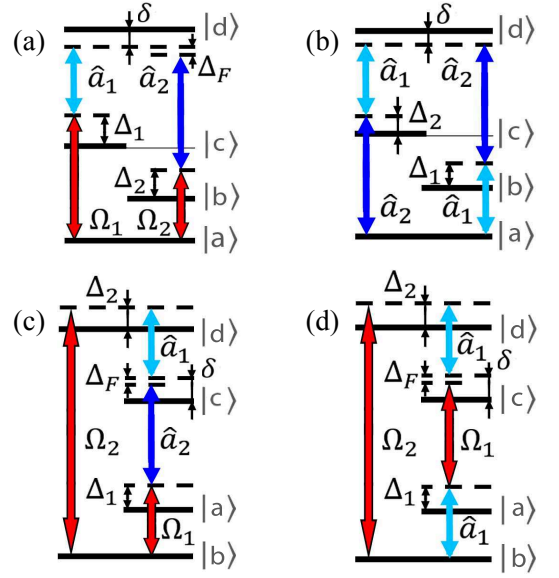}%
\caption{(Color online) Energy diagram for the FWM schemes of (a) beam-splitter operation, (b) cross-Kerr interaction, (c) two-mode squeezing, and (d) single-mode squeezing. The labels are: detuning $\Delta_{i}$ for single-photon transition, detuning $\delta$ for two-photon process, detuning $\Delta_{F}$ for four-photon process, Rabi frequency $\Omega_{i}$ for classical driving (red arrows), and operator $\hat{a}_{i}$ for resonator mode (light and dark blue arrows).}%
\label{fig4_FWM}%
\end{figure}

In this section, we present the realization of four quantum operations described in Appendix A using the quantum toolbox. The system parameters are chosen as the following: $\omega _{a_{1}}/2\pi= 10\,\textrm{GHz}$ and $\omega_{a_{2}}/2\pi=16\,\textrm{GHz}$ for the resonator frequencies, $g_{1}/2\pi=g_{2}/2\pi=0.3\,\textrm{GHz}$ for the coupling constants, and $E_{mx}/2\pi\hbar=4\,\textrm{GHz}$ for the capacitive coupling defined in Eq.~(\ref{Emx}). The effective Josephson energies $E_{J1}$, $E_{J2}$, and $E_{Jm}$ can be adjusted by tuning the magnetic flux $\Phi_{1}$, $\Phi_{2}$, and $\Phi$ respectively. The energy diagrams of the dispersive FWM schemes for the operations are shown in Fig.~\ref{fig4_FWM}.

\subsection{Beam-splitter operation\label{beamsplitter}}
Consider the energy diagram in Fig.~\ref{fig4_FWM}a, where the detunings are defined as $\Delta_{1}=\omega_{1}-E_{ca}/\hbar$, $\Delta_{2}=\omega_{2}-E_{ba}/\hbar$, and $\delta=\omega_{1}+\omega_{a_{1}}-E_{da}/\hbar$ with $\omega_{i}$ being the frequency of the classical driving. In the energy diagram, the resonator modes only couple with the upper transitions $\sigma_{db},\,\sigma_{dc}$ and the classical drivings only couple with the lower transitions $\sigma_{ab},\,\sigma_{ac}$. This can be achieved by choosing the parameters $b_{0}= -0.68$, $E_{J1}/2\pi\hbar=8\,\textrm{GHz}$, and $E_{J2}/2\pi\hbar=15\,\textrm{GHz}$ so that the large separation between the ground state and the excited states provides energy selection for the transitions. Under these parameters and appropriate classical driving frequencies, the detunings for the desired single-photon and two-photon processes as labelled in the energy diagram are $\Delta_{i}/2\pi=-4\,\textrm{GHz}$ and $\delta/2\pi=3.49\,\textrm{GHz}$. The detunings for the unwanted transitions, e.g. $\omega_{a_{1}}-E_{ba}/\hbar,  \omega_{a_{2}}-E_{ca}/\hbar, \sim 9\,\textrm{GHz}$, are much larger so that the unwanted transitions are suppressed. These parameters also give $|\cos(\theta_{+}-\theta_{-})|\gg|\sin(\theta_{+}-\theta_{-})|$ so that the coupling between the mode $\hat{a}_{1}$ ($\hat{a}_{2}$) and the transition $\sigma_{db}$ ($\sigma_{dc}$) is much weaker than the coupling between the mode $\hat{a}_{1}$ ($\hat{a}_{2}$) and the transition $\sigma_{dc}$ ($\sigma_{db}$). As a result, the mode $\hat{a}_{1}$ mainly couples with the $\sigma_{dc}$ transition. Similar arguments apply to other transitions in the energy diagram. 

We divide the total Hamiltonian into two parts: $H_{tot}=H_{0}+V$ where 
\begin{equation}
H_{0}/\hbar = \omega_{1} \sigma_{cc}+\omega_{2}\sigma _{bb}+(\omega_{1}+\omega_{a_{1}}) \sigma _{dd}+\sum_{i}\omega_{a_{i}}\hat{a}_{i}^{\dagger}\hat{a}_{i},\label{H0BS} 
\end{equation}
$\sigma_{ii}=|i\rangle\langle i|$, and $V$ includes all remaining terms in the total Hamiltonian. Here, the ground state energy is set to zero. The Hamiltonian in the interaction picture of $H_{0}$ can be written as $e^{iH_{0}t/\hbar}Ve^{-iH_{0}t/\hbar}\approx H_{I0}+V_{I}$ under the rotating wave approximation with $H_{I0}/\hbar=-\Delta _{1}\sigma_{cc}-\Delta _{2}\sigma _{bb}-\delta \sigma _{dd}$ and
\begin{equation}
V_{I}/\hbar = \Omega _{1}\sigma _{ca}+\Omega _{2}\sigma _{ba} + \tilde{g}_{1}\hat{a}_{1}\sigma _{dc}+\tilde{g}_{2}e^{i\Delta_{F}t}\hat{a}_{2}\sigma_{db}+h.c.. \label{HintBS}
\end{equation}
The effective couplings $\tilde{g}_{1}=g_{1}\cos(\theta_{+}-\theta_{-})$ and $\tilde{g}_{2}=g_{2}\cos(\theta_{+}+\theta_{-})$ are derived from Eqs.~(\ref{sigmaX1}, \ref{sigmaX2}). And, $\Delta _{F} = \omega _{1}+\omega _{a_{1}}-\omega _{2}-\omega _{a_{2}} $ is a small detuning for the four-photon process designed to balance the extra terms in the effective Hamiltonian for the resonators.

Given the dispersive conditions discussed in Sec.~\ref{sec3}, we treat $V_{I}$ as a perturbation to the Hamiltonian $H_{I0}$. Assume that the toolbox is initially prepared in its ground state $|a\rangle$. It can be shown that the dominant correction to $H_{I0}$ by the perturbation $V_{I}$ is a fourth-order term that generates a ``quantum'' energy shift in the ground state in the form of $\sigma_{aa} H_{\textrm{bm}}$ where 
\begin{equation}
H_{\textrm{bm}}/\hbar=\sum_{i}\delta \epsilon _{i}^{\textrm{bm}}\hat{a}_{i}^{\dag }\hat{a}_{i}+\left( \chi^{\textrm{bm}}\hat{a}_{1}^{\dag }\hat{a}_{2}e^{i\Delta _{F}t}+h.c.\right)\label{EffBS1}
\end{equation}
with the energy shifts $\delta \epsilon _{i}^{\textrm{bm}}=\Omega _{i}^{2}\tilde{g}_{i}^{2}/\Delta_{i}^{2}\delta $, the effective coupling constant
\begin{equation}
\chi^{\textrm{bm}}=\Omega_{1}\Omega_{2}\tilde{g}_{1}\tilde{g}_{2}/\Delta _{1}\Delta _{2}\delta, \label{BS1coupling}
\end{equation}
and the four-photon detuning $\Delta_{F}=\delta \epsilon_{2}^{\textrm{bm}}-\delta \epsilon_{1}^{\textrm{bm}}$. The small four-photon detuning is chosen to balance the effect of the mode shifts $\delta \epsilon _{i}^{\textrm{bm}}$. The above effective Hamiltonian performs the beam-splitter operation on the resonators while the toolbox is preserved in the ground state during the operation. The beam-splitter operation can perform a swap gate on the resonators after applied for a gate time $t_{g}=\pi/2\vert \chi^{\textrm{bm}}\vert$.\cite{KokRMP2007}  With the parameters given above and with $\Omega_{i}/2\pi=1.5\,\textrm{GHz}$, the effective coupling is $|\chi^{\textrm{bm}}|/2\pi=6.2\,\textrm{MHz}$ and the swap gate has a gate time of $40.5\,\textrm{ns}$.  

\subsection{Cross-Kerr nonlinearity\label{crossKerr}}
The energy diagram for the cross-Kerr operation is shown in Fig.~\ref{fig4_FWM}b with the detunings defined as $\Delta_{1}=\omega_{a_{1}}-E_{ba}/\hbar$, $\Delta_{2}=\omega_{a_{2}}-E_{ca}/\hbar$, and $\delta=\omega_{a_{1}}+\omega_{a_{2}}-E_{da}/\hbar$. Compared with the beam-splitter operation, the asymmetry in the energy levels is reduced and no classical driving needs to be applied. The parameters are chosen to be $b_{0}=-0.61$, $E_{J1}/2\pi\hbar=8.45\,\textrm{GHz}$, and $E_{J2}/2\pi\hbar=13.95\,\textrm{GHz}$, which give $\Delta_{1}/2\pi=-4.59\,\textrm{GHz}$, $\Delta_{2}/2\pi=-4.93\,\textrm{GHz}$, and $\delta/2\pi=0.17\,\textrm{GHz}$. Under these parameters, the effective coupling constants for the desired couplings $\hat{a}_{1}\sigma_{dc}, \hat{a}_{1}\sigma_{ab}$ are $\tilde{g}_{1}/2\pi\approx 0.3\,\textrm{GHz}$. While the unwanted couplings $\hat{a}_{1}\sigma_{db}, \hat{a}_{1}\sigma_{ac}$ are strongly suppressed with $\tilde{g}_{1}/2\pi\approx 0.01\,\textrm{GHz}$. The unwanted coupling $\hat{a}_{2} \sigma_{dc}$ has a detuning of $\vert\omega_{a_{2}}-E_{dc}/\hbar\vert/2\pi\approx 11\,\textrm{GHz}$ and is suppressed by the large detuning. 

We divide the total Hamiltonian as $H_{tot}=H_{0}+V$ with
\begin{equation}
H_{0}/\hbar = \omega_{a_{1}} \sigma_{bb}+\omega_{a_{2}}\sigma _{cc}+\omega_{t} \sigma _{dd}+\sum_{i}\omega_{a_{i}}\hat{a}_{i}^{\dagger}\hat{a}_{i}\label{H0CK}
\end{equation}
and $\omega_{t}=\omega_{a_{1}}+\omega_{a_{2}}$. The Hamiltonian in the interaction picture of $H_{0}$ is $e^{iH_{0}t/\hbar}Ve^{-iH_{0}t/\hbar}\approx H_{I0}+V_{I}$ under the rotating wave approximation where $H_{I0}/\hbar=-\Delta _{1}\sigma_{bb}-\Delta _{2}\sigma _{cc}-\delta \sigma _{dd}$ and 
\begin{equation}
V_{I}/\hbar = \tilde{g}_{1}\hat{a}_{1}(\sigma _{dc}+\sigma_{ba})+\tilde{g}_{2}\hat{a}_{2}(\sigma _{db}+\sigma_{ca})+h.c. \label{HintCK}
\end{equation}
with the effective couplings $\tilde{g}_{1}=g_{1}\cos(\theta_{+}-\theta_{-})$ and $\tilde{g}_{2}=g_{2}\cos(\theta_{+}+\theta_{-})$. With the same perturbation theory approach as used in the previous subsection, we derive the fourth-order perturbation correction to the ground state energy $\sigma_{aa} H_{\textrm{ck}}$. Here,
\begin{equation}
H_{\textrm{ck}}/\hbar=\sum_{i}\delta \epsilon_{i}^{\textrm{ck}}\hat{a}_{i}^{\dag }\hat{a}_{i}+\chi^{\textrm{ck}}\hat{a}_{1}^{\dag }\hat{a}_{2}^{\dag}\hat{a}_{2}\hat{a}_{1}, \label{EffCK}
\end{equation}
with the energy shifts $\delta \epsilon_{i}^{\textrm{ck}}=\tilde{g}_{i}^{2}/\Delta _{i}$ and the effective coupling
\begin{equation}
\chi^{\textrm{ck}}=\left(1/\Delta_{1}+1/\Delta_{2}\right)^{2}\left(\tilde{g}_{1}^{2}\tilde{g}_{2}^{2}/\delta\right).\label{CKcoupling}
\end{equation}
With the above parameters, $|\chi^{\textrm{ck}}|/2\pi=6.3\,\textrm{MHz}$.

\begin{figure}[ptb]
\includegraphics[width=7.5cm,clip]{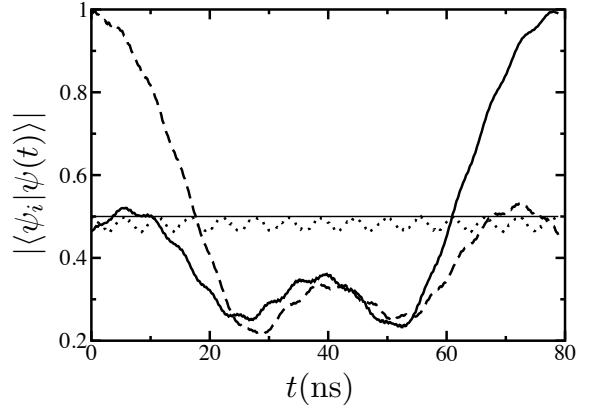}%
\caption{Time evolution of $\vert\langle \psi_{i}|\psi(t)\rangle\vert$ under the cross-Kerr interaction. Solid curve: $\vert\psi_{i}\rangle$ is the target state at time $t_{g}$; dashed curve: $\vert\psi_{i}\rangle$ is the initial state given in Eq.~(\ref{initial_state}); thin-solid curve: $\vert\psi_{i}\rangle=|a\rangle|0_{1}0_{2}\rangle$; and dotted curve: $\vert\psi_{i}\rangle=|a\rangle|1_{1}1_{2}\rangle$.}
\label{fig5_simu_ck}%
\end{figure}
Given the effective cross-Kerr interaction in Eq.~(\ref{EffCK}), the controlled phase gate can be realized on the resonator modes which generates the following transformations
\begin{subequations}
\begin{align}
|0_{1}0_{2}\rangle & \rightarrow |0_{1}0_{2}\rangle,\label{gate00}\\
|0_{1}1_{2}\rangle & \rightarrow e^{-i\delta \epsilon_{2}^{\textrm{ck}} t} |0_{1}1_{2}\rangle,\label{gate01}\\
|1_{1}0_{2}\rangle & \rightarrow e^{-i\delta \epsilon_{1}^{\textrm{ck}} t} |1_{1}0_{2}\rangle,\label{gate10}\\
|1_{1}1_{2}\rangle & \rightarrow e^{-i\delta \epsilon_{1}^{\textrm{ck}} t-i\delta \epsilon_{2}^{\textrm{ck}} t-i \chi^{\textrm{ck}} t}  |1_{1}1_{2}\rangle,\label{gate11}
\end{align}
\end{subequations}
on the corresponding resonator states. At $t_{g}=\pi/|\chi^{\textrm{ck}}|$, the nonlinear term generates a phase $\pi$ on the state $|1_{1}1_{2}\rangle$. With the above parameters, we find that the gate time is $t_{g}=79.4\,\textrm{ns}$. To test the above results, we simulate the time evolution of the combined resonator-toolbox system under the Hamiltonian $H_{I0}+V_{I}$ given in Eq.~(\ref{HintCK}). Consider the initial state
\begin{equation}
|a\rangle (|0_{1}\rangle+|1_{1}\rangle)(|0_{2}\rangle+|1_{2}\rangle)/2,\label{initial_state}
\end{equation}
where the toolbox is in the ground state $|a\rangle$. The target state at time $t_{g}$ can be derived by substituting time $t$ with $t_{g}$ in Eqs.~(\ref{gate00},\ref{gate01},\ref{gate10},\ref{gate11}) while keeping the toolbox in the state $|a\rangle$. In Fig.~\ref{fig5_simu_ck}, we plot $\vert\langle \psi_{i}|\psi(t)\rangle\vert$ for different states $\vert\psi_{i}\rangle$. After the evolution starts, the amplitude of the initial state starts to decrease. At the target time $t_{g}$, the amplitude of the target state nearly reaches unity, which demonstrates that the final state of the system has successfully evolved to become the target state. The amplitude of the state $|a\rangle|0_{1}0_{2}\rangle$ stays at $0.5$ with no significant transition to other states during the evolution. Meanwhile, the amplitude of the state $|a\rangle|1_{1}1_{2}\rangle$ shows small deviation from $0.5$ due to leakage to the excited states of the toolbox during the evolution, which can reduce the fidelity of the controlled phase gate. 

\subsection{Two-mode squeezing\label{2modesqueezing}}
In Fig.~\ref{fig4_FWM}c, we present an energy diagram that can realize the two-mode squeezing operation with the detunings defined as $\Delta_{1}=\omega_{1}-E_{ab}/\hbar$, $\Delta_{2}=\omega_{2}-E_{db}/\hbar$, and $\delta=\omega_{2}-\omega_{a_{1}}-E_{cb}/\hbar$. Here, the order of the eigenstates is switched with $E_{a}>E_{b}$, i.e. the state $|b\rangle$ is now the ground state of the toolbox. The circuit parameters are chosen to be $b_{0}=1.2$, $E_{J1}/2\pi\hbar=8\,\textrm{GHz}$ and $E_{J2}/2\pi\hbar=14\,\textrm{GHz}$. Under these parameters, we tune the classical driving frequencies to give $\Delta_{1}/2\pi=3\,\textrm{GHz}$, $\Delta_{2}/2\pi=-5\,\textrm{GHz}$ and $\delta/2\pi=-3.67\,\textrm{GHz}$. The classical drivings induce the $\sigma_{ab}, \sigma_{db}$ transitions in the energy diagram. It can also be shown that the mode $\hat{a}_{1}$ mainly couples with the $\sigma_{dc}$ transition and the mode $\hat{a}_{2}$ mainly couples with the $\sigma _{ac}$ transition. With these parameters, the driving frequencies are $\omega_{1}/2\pi=10.9\,\textrm{GHz}$ and $\omega_{2}/2\pi=24.9\,\textrm{GHz}$. Note that at high driving frequency, the microwave driving can induce quasiparticle excitations in the superconductor. This effect can be avoided by considering a two-photon process for the $\sigma _{db}$ transition.

Using the approach in the previous subsections with 
\begin{equation}
H_{0}/\hbar = \omega_{1} \sigma_{aa}+\omega_{s} \sigma _{cc}+\omega_{2}\sigma _{dd}+\sum_{i}\omega_{a_{i}}\hat{a}_{i}^{\dagger}\hat{a}_{i},\label{H0SQ2}
\end{equation}
and $\omega_{s}=\omega_{2}-\omega_{a_{1}}$, we derive that $H_{I0}/\hbar=-\Delta _{1}\sigma_{aa}-\Delta _{2}\sigma _{dd}-\delta \sigma _{cc}$ and 
\begin{equation}
V_I/\hbar=\Omega _{1}\sigma _{ab}+\Omega _{2}\sigma _{db}+\tilde{g}_{1}\hat{a}_{1}\sigma _{dc}+\tilde{g}_{2}e^{i\Delta_{F}t}\hat{a}_{2}\sigma_{ac}+h.c.  \label{HintSQ2}
\end{equation}
with the effective couplings $\tilde{g}_{1}=g_{1}\cos(\theta_{+}-\theta_{-})$ and $\tilde{g}_{2}=g_{2}\cos(\theta_{+}+\theta_{-})$, and a small four-photon detuning $\Delta _{F}=\omega _{2}-\omega _{1}-\omega _{a_{1}}-\omega _{a_{2}}$. The fourth-order perturbation correction to the ground state energy is $\sigma_{aa} H_{\textrm{sq}}$ with
\begin{equation}
H_{\textrm{sq}}/\hbar=\sum_{i}\delta \epsilon _{i}^{\textrm{sq}}\hat{a}_{i}^{\dag }\hat{a}_{i}+\left( \chi^{\textrm{sq}}\hat{a}_{1}^{\dag }\hat{a}_{2}^{\dag}e^{i\Delta_{F}t}+h.c.\right),\label{EffSQ2}
\end{equation}
where the energy shifts are $\delta \epsilon _{i}^{\textrm{sq}}=\Omega _{\bar{i}}^{2}\tilde{g}_{i}^{2}/\Delta_{\bar{i}}^{2}\delta$ with the index $\bar{i}$ referring to the circuit elements in the opposite side of the $i$th resonator (e.g. for $i=1$, $\bar{i}=2$), the effective coupling is 
\begin{equation}
\chi^{\textrm{sq}}=\Omega _{1}\Omega _{2}\tilde{g}_{1}\tilde{g}_{2}/\Delta _{1}\Delta_{2}\delta,\label{SQ2coupling}
\end{equation}
and the four-photon detuning is $\Delta_{F}=-\delta \epsilon_{1}^{\textrm{sq}}-\delta \epsilon_{2}^{\textrm{sq}}$. The nonzero four-photon detuning balances the effect of the energy shifts and makes the squeezing operation possible. With the parameters given above and with $\Omega_{i}/2\pi=2\,\textrm{GHz}$, we have $|\chi^{\textrm{sq}}|/2\pi=4.3\,\textrm{MHz}$.

\subsection{Single-mode squeezing\label{1modesqueezing}}
The single-mode squeezing operation can be realized with the energy diagram in Fig.~\ref{fig4_FWM}d with the detunings defined as $\Delta_{1}=\omega_{a_{1}}-E_{ab}/\hbar$, $\Delta_{2}=\omega_{2}-E_{db}/\hbar$, and $\delta=\omega_{2}-\omega_{a_{1}}-E_{cb}/\hbar$. The circuit parameters are chosen to be $b_{0}=1.2$, $E_{J1}/2\pi\hbar=6\,\textrm{GHz}$ and $E_{J2}/2\pi\hbar=15\,\textrm{GHz}$. Tuning the frequencies of the classical drivings, we have $\Delta_{1}/2\pi=3\,\textrm{GHz}$, $\Delta_{2}/2\pi=-5\,\textrm{GHz}$, and $\delta/2\pi=-4.75\,\textrm{GHz}$. The circuit parameters are chosen to have the energy separation $E_{ab}$ close to the frequency of mode $\hat{a}_{1}$. Hence, $\hat{a}_{1}$ couples strongly with the $\sigma_{ab}$ transition as well as the $\sigma_{dc}$ transition. Meanwhile, the energy levels are adjusted so that the frequency of mode $\hat{a}_{2}$ is largely detuned from all possible transitions and is effectively decoupled from the toolbox. The classical drivings generate the $\sigma_{ac}, \sigma_{db}$ transitions. With the above parameters, $\omega_{1}/2\pi=12.0\,\textrm{GHz}$ and $\omega_{2}/2\pi=25.0\,\textrm{GHz}$. 

With the Hamiltonian
\begin{equation}
H_{0}/\hbar = \omega_{a_{1}} \sigma_{aa}+\omega_{s} \sigma _{cc}+\omega_{2}\sigma _{dd}+\sum_{i}\omega_{a_{i}}\hat{a}_{i}^{\dagger}\hat{a}_{i},\label{H0SQ1}
\end{equation}
and $\omega_{s}=\omega_{2}-\omega_{a_{1}}$, we have $H_{I0}/\hbar=-\Delta _{1}\sigma_{aa}-\Delta _{2}\sigma _{dd}-\delta \sigma _{cc}$ and 
\begin{equation}
V_I/\hbar=\Omega _{1}\sigma _{ca}e^{i\Delta_{F}t}+\Omega _{2}\sigma _{db}+\tilde{g}_{1}\hat{a}_{1}(\sigma _{dc}+\sigma_{ab})+h.c.  \label{HintSQ1}
\end{equation}
with the effective coupling constant $\tilde{g}_{1}=g_{1}\cos(\theta_{+}-\theta_{-})$ and the four photon detuning $\Delta _{F}=\omega _{2}-\omega _{1}-2\omega _{a_{1}}$. Then, we derive the fourth-order perturbation correction to the ground state energy $\sigma_{aa} H_{\textrm{sq1}}$ with
\begin{equation}
H_{\textrm{sq1}}/\hbar=\delta \epsilon_{1}^{\textrm{sq1}}\hat{a}_{1}^{\dag }\hat{a}_{1}+\left( \chi^{\textrm{sq1}}\hat{a}_{1}^{\dag }\hat{a}_{1}^{\dag}e^{-i\Delta
_{F}t}+h.c.\right)\label{EffSQ1}
\end{equation}
where the energy shifts are
\begin{equation}
\delta \epsilon _{1}^{\textrm{sq1}}=\left( \delta /\Delta _{1}+\Omega_{1}^{2}/\Delta _{1}^{2}+\Omega _{2}^{2}/\Delta _{2}^{2}\right) \left(\tilde{g}_{1}^{2}/\delta\right), \label{eq:dep}
\end{equation}
the effective coupling  constant is
\begin{equation}
\chi^{\textrm{sq1}}=\Omega _{1}\Omega _{2}\tilde{g}_{1}^{2}/\Delta _{1}\Delta_{2}\delta, \label{SQ1coupling}
\end{equation}
and the four-photon detuning is $\Delta_{F}=2\delta \epsilon_{1}^{\textrm{sq1}}$. With the parameters given above and with $\Omega_{i}/2\pi=1\,\textrm{GHz}$, we have $|\chi^{\textrm{sq1}}|/2\pi=11.1\,\textrm{MHz}$.

\section{error sources\label{sec5}}
Quantum errors can affect the effective quantum operations on the superconducting resonators. In our system, the main sources of quantum errors include (1) unwanted transitions induced by resonator-qubit coupling and classical driving, (2) indirect coupling due to the cross-talk between different circuit elements, and (3) decoherence of the qubits and resonators.  

\begin{figure}[ptb]
\includegraphics[width=7cm,clip]{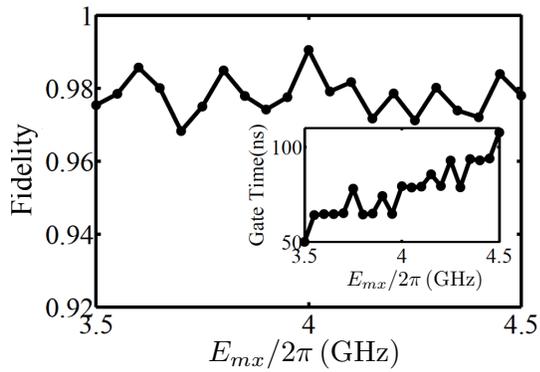}%
\caption{Fidelity and gate time versus $E_{mx}$ for the controlled phase gate using parameters in Sec.~\ref{crossKerr}.}%
\label{fig6_simulation}%
\end{figure}
We first study the effect of unwanted transitions on the effective quantum operations studied in Sec.~\ref{sec4}. In the proposed scheme, the circuit parameters can be adjusted to suppress the unwanted transitions either by reducing the coupling matrix element of the unwanted transitions or by varying the energy levels of the unwanted transitions to produce large detuning. To test the effectiveness of our approach,  we numerically simulate the quantum operations using the full Hamiltonian in Eq.~(\ref{Hexact}) which includes all of the nonzero transition matrix elements. In Fig.~\ref{fig6_simulation}, we present our result for the fidelity of the controlled phase gate which is demonstrated in Sec.~\ref{crossKerr}. The initial state for the simulation is given in Eq.~(\ref{initial_state}). For each value of the coupling energy $E_{mx}$ with $E_{mx}/2\pi \in (3.5, 4.5)\,\textrm{GHz}$, we search for the maximum fidelity by varying $E_{Ji}$, $b_{0}$, and the gate time. The maximum fidelity and the corresponding gate time are then plotted in Fig.~\ref{fig6_simulation}. We show that the maximum fidelity can exceed $0.99$ at $E_{mx}/2\pi=4\,\textrm{GHz}$. Our simulation hence shows that high fidelity can be achieved for effective quantum operations by choosing appropriate parameters. In Fig.~\ref{fig6_simulation}, the fidelity fluctuates ``randomly'' within a narrow range as $E_{mx}$ varies. This is because the time evolution of the total system includes small but fast oscillations resulted from multiple unwanted off-resonant transitions, as can be seen in Fig.~\ref{fig5_simu_ck}. Such oscillations affect the optimal gate time and the fidelity by a small magnitude in a nearly ``random'' way. 

Another source of quantum errors is the circuit cross-talk between different circuit elements. In Eqs.~(\ref{Hr}, \ref{coupling}), the direct coupling between the resonator and its neighboring qubit is presented. However, due to the cross-talk, each resonator also couples with the qubit in the opposite side of the circuit in the form of $\hbar g_{i}^{(2)}\sigma_{xi} (\hat{a}^{\dagger}_{\bar{i}}+\hat{a}_{\bar{i}})$. The coupling constant for this indirect coupling is
\begin{equation}
g_{i}^{(2)}=\left(  \frac{C_{gi}C_{m}}{C_{\Sigma 1}C_{\Sigma 2}-C_{m}^{2}}\right)  \left(\frac{e^{2}}{2C_{\Sigma ri}}\hbar\omega_{a_{i}}\right)^{1/2} \label{gi2}
\end{equation}
with $g_{i}^{(2)}/g_{i}=C_{m}/C_{\Sigma \bar{i}}\ll 1$.  Another indirect coupling is the coupling between the two resonators in the form of $g^{(3)}(a_{1}+a_{1}^{\dag})(a_{2}+a_{2}^{\dag})$ with the coupling constant
\begin{equation}
g^{(3)}= \frac{\sqrt{C_{g1}C_{g2}}C_{m}}{C_{\Sigma 1}C_{\Sigma 2}-C_{m}^{2}}  \sqrt{  \frac{C_{g1}C_{g2}}{4 C_{\Sigma r1}C_{\Sigma r2}}} \sqrt{\hbar\omega_{a_{1}}\hbar\omega_{a_{2}}}.
\end{equation}
It can be shown that 
\begin{equation}
g^{(3)}/g_{i}\sim \frac{C_{m}}{C_{\Sigma \bar{i}}}\sqrt{\frac{C_{gi}}{4 C_{\Sigma ri}}}\sqrt{\frac{\hbar\omega_{a_{i}}}{e^{2}/2C_{gi}}}\ll 1.
\end{equation}
These indirect coupling terms can hence be neglected.

Decoherence of the superconducting qubits and resonators is one of the key barriers for scalable quantum information processing and has been intensively studied in the past two decades.\cite{HPaikSchoelkopfPRL2011, RigettiSteffenPreprint2012} The quantum toolbox studied here is made of two coupled superconducting qubits and can be subject to both single-qubit and two-qubit decoherence. However, for the dispersive FWM scheme proposed in this work, the quantum toolbox is largely preserved in its ground state during the quantum operations, and hence, is not affected by either single-qubit or two-qubit decoherence. Only small leakage to the excited states can be induced by the unwanted transitions as is studied above. Given the small amplitude of the leakage as is shown in Fig.~\ref{fig5_simu_ck}, qubit decoherence will not affect the effective quantum operations significantly. Superconducting resonators can have relatively high Q-factors. With $Q$ exceeding $10^{6}$, it corresponds to a damping time of $\sim 100\,\mu\textrm{s}$. Meanwhile, our study in the previous section shows that the time scale for the quantum operations - $\sim \pi/\vert\chi^{\alpha}\vert$ for operation $\alpha$ - is below $100\,\textrm{ns}$ which is shorter than the decoherence time by $2 - 3$ orders of magnitude. 

\section{conclusions\label{sec6}}
To conclude, we presented a superconducting quantum toolbox that can perform various quantum operations on superconducting resonators in one single circuit. The scheme exploits the dispersive FWM approach to generate effective couplings between the resonator modes. By adjusting the circuit parameters, the energy levels and coupling constants can be varied to generate specific quantum operation. We discussed the main error sources in the schemes and numerically simulated the controlled phase gate on Fock states. Our results showed that high-fidelity quantum operations can be achieved in this circuit. One advantage of this scheme is that nearly all quantum operations for both discrete-state and continuous-variable quantum protocols can be realized in one single circuit.\cite{KokRMP2007, LloydBraunsteinPRL1998, ShihBook} Our scheme can advance the implementation of quantum information processing on the microwave modes in the superconducting circuits.

\section*{ACKNOWLEDGEMENTS}
We would like to thank Prof. A. N. Korotkov and Prof. A. Lupa\c{s}cu for valuable comments on the manuscript. This work is supported by the projects NSF-CCF-0916303 and NSF-DMR-0956064. XHD is partially supported by a CSC Scholarship.

\section*{APPENDIX A\label{appen}}
Quantum information processing with photons has been widely studied either with discrete states or continuous variable states.\cite{KokRMP2007, LloydBraunsteinPRL1998} Here, we briefly summarize the basic quantum operations for the photon modes.\cite{ShihBook} Two categories of quantum operations are considered: the Bogoliubov-linear operations and the nonlinear interactions. The photon modes are represented by the annihilation (creation) operators $\hat{a}_{1}$ and $\hat{a}_{2}$ ($\hat{a}^{\dag}_{1}$ and $\hat{a}^{\dag}_{2}$).

\subsection{Bogoliubov-linear operations}
The Bogoliubov-linear operations perform the following transformation 
\begin{equation}
\hat{a}_{i}\rightarrow \sum_{j} A_{ij}\hat{a}_{j} + B_{ij}\hat{a}_{j}^{\dag} + C_{i}\label{trans}
\end{equation}
with coefficients $A_{ij},\,B_{ij}\,C_{i}$.  An arbitrary Bogoliubov-linear operation can be constructed using the basic elements: the beam-splitter operation, the squeezing operation, and the phase shifter. 

The beam-splitter operation can be realized by $H_{\textrm{bm}}=\hbar \chi^{\textrm{bm}} e^{i\phi}\hat{a}^{\dagger}_{1}\hat{a}_{2}+h.c.$ with coupling amplitude $\chi^{\textrm{bm}}$. Under this Hamiltonian, the operators evolve as
\begin{equation}
\left( \begin{array} [c]{c}%
\hat{a}_{1}(t)\\ \hat{a}_{2}(t)%
\end{array} \right)  =\left( 
\begin{array} [c]{cc}%
\cos\varphi & -e^{i\phi}\sin\varphi\\ -e^{-i\phi}\sin\varphi& \cos\varphi %
\end{array} \right)  \left( \begin{array} [c]{c}%
\hat{a}_{1}(0)\\ \hat{a}_{2}(0)%
\end{array} \right)  
\label{idealBMHeis}%
\end{equation}
with $\varphi=\chi_{\textrm{bm}}t $. At $\varphi=\pi/2$, the beam-splitter operation swaps the states of the two modes up to a phase factor. In discrete-state quantum computing schemes, this operation can generate single-qubit Hadamard gate.  

The squeezing operation can be realized by $H_{\textrm{sq}}=i \hbar \chi^{\textrm{sq}}\hat{a}^{\dagger}_{1}\hat{a}^{\dagger}_{2}+h.c.$ with coupling amplitude $\chi^{\textrm{sq}}$. Under this Hamiltonian, the operators evolve as
\begin{equation}
\left( \begin{array} [c]{c}%
\hat{a}_{1}(t)\\ \hat{a}^{\dag}_{2}(t)%
\end{array} \right)  = \left( \begin{array}[c]{cc}%
\cosh\varphi & \sinh\varphi\\ \sinh\varphi& \cosh\varphi %
\end{array} \right)  \left(\begin{array} [c]{c} %
\hat{a}_{1}(0)\\ \hat{a}^{\dag}_{2}(0)%
\end{array}\right)  
\label{idealSQHeis}%
\end{equation}
with $\varphi=\chi^{\textrm{sq}}t $, which describes the parametric amplification process that generates two-mode squeezing.\cite{ShihBook} When applied to the vacuum state, it generates the so-called continuous-variable EPR states. When combined with the beam-splitter operation, it can generate squeezing on individual mode. A related operation is the single-mode squeezing operation which can be generated by $H_{\textrm{sq1}}=i \hbar \chi^{\textrm{sq1}}(\hat{a}^{\dagger})_{i}^{2}+h.c.$ on mode $\hat{a}_{i}$.

The phase shifter operation can be realized by $H_{\textrm{ph}}=\hbar \Delta^{\textrm{ph}}\hat{a}^{\dag}_{i}\hat{a}_{i}$ which creates a shift in the resonator frequency. The above quantum operations can be combined to generate arbitrary linear transformations in Eq.~(\ref{trans}).

\subsection{Cross-Kerr nonlinearity}
One nonlinear operation is the cross-Kerr interaction given by $H_{\textrm{ck}}=\hbar \chi^{\textrm{ck}} \hat{a}^{\dagger}_{1}\hat{a}_{1}\hat{a}^{\dagger}_{2}\hat{a}_{2}$ between two modes with interaction amplitude $\chi^{\textrm{ck}}$. This interaction can lead to controlled gates on photon qubits.\cite{KokRMP2007} For continuous-variable schemes, this operation together with the linear operations can generate operations that are arbitrary polynomials of the quadrature variables.\cite{BoydBook, KokRMP2007, LloydBraunsteinPRL1998} This operation can also be exploited for quantum nondemolition measurement on photon states.

\end{document}